\documentclass[peerreview, 12pt]{IEEEtran}

\usepackage{cite}

\usepackage{graphicx}

\usepackage{url}
\usepackage{amsmath}
\usepackage{amsfonts}
\usepackage{amssymb}
\usepackage[pdftex,hypertexnames=false]{hyperref}

\begin{document}
\title{\Large{Performance Analysis of Cooperative Multi-Carrier Relay-Based UAV Networks Over Generalized Fading Channels}}
%
%
%
%

\author{\small{\IEEEauthorblockN{$^{(1)}$Ibrahim Y. Abualhaol and $^{(2)}$Mustafa M. Matalgah}\\
\IEEEauthorblockA{ $^{(1)}$Khalifa University, Sharjah, United Arab Emirates, \\$^{(2)}$The University of Mississippi, University, MS 38677\\
Email: $^{(1)}$ibrahimee@ieee.org, $^{(2)}$mustafa@olemiss.edu}}}

\maketitle
\begin{abstract}

The outage probability in a network of cooperative unmanned airborne
vehicles (UAVs) over generalized fading channels is studied
analytically using finite mixture with expectation maximization
technique. A relay-based topology with one ground control unit (GCU)
is considered, where the cooperative UAVs can communicate with the
GCU directly or through relay. The application the UAV assigned for,
specifies the minimum required transmission rate the UAV should
achieve. The outage probability of the system is defined as the
probability that either the transmission rate over any of the links
drops below a predefined minimum threshold for that link or the
Relay-GCU link is not able to transmit the aggregate data from all
relayed UAVs and the minimum rate required by the relay UAV itself.
Throughout the paper, expressions for the outage probability and the
average achievable bit rate of a cooperative multi-carrier system
are derived over generalized fading channels. Finite Mixture with
Expectation-Maximization algorithm is utilized to derive a simple
approximate expression for the probability density function (pdf) of
the achievable bit rate assuming adaptive M-ary quadrature amplitude
modulation (M-QAM). This pdf is used to derive closed-form
expressions for the outage probability and the average bit rate.
\end{abstract}

\begin{IEEEkeywords}
Outage probability, cooperative UAVs, fading channels,
Finite-Mixture, relay-based, adaptive modulation, multi-carrier
communications.
\end{IEEEkeywords}

\newpage
\section{Introduction}
Unmanned airborne vehicles (UAVs) have evolved into high-tech
capable small vehicles, used by the armed forces worldwide, mostly
for surveillance and data acquisition purposes. UAVs have also been
used in many civilian applications such as agricultural purposes,
natural resources management, and natural disaster response. The
demand for these products in the commercial industry arises from the
low manufacturing and operational costs of the systems, the
flexibility of the aircrafts to adjust to the particular needs of
the consumer, and the elimination of the risk of human lives
(pilots) in applications that require difficult missions
\cite{UAV_Book_1} and \cite{UAV_Book_2}. The communication links in
a swarm of UAVs suffer from many problems, one of which is the power
fluctuation of the received signal due to multi-path propagation and
Doppler spread, which becomes more severe at high speeds and high
carrier frequencies \cite{UAV_Book_3}. One method to combat this
problem, is the use of multi-carrier system with good resource
management, which results in increased reliability, bandwidth
efficiency, and power efficiency of the communication system
~\cite{combinedDSSSFHSS}. ~A more efficient method is to integrate
multi-carrier communication systems with relay-based cooperative
techniques, which is a one feasible solution to overcome the
severity of the UAVs communication link.

The first formulations of general relaying problems appeared in the
information theory community in \cite{Capacity_Theorems} and
\cite{Three_Terminals_71}. The traditional relay channel model is
comprised of three nodes: a source that transmits information, a
destination  that receives information, and a relay  that both
receives and transmits information to enhance communication between
the source and the destination. Recently, many new models with
multiple relays have been proposed and examined (see e.g.,
\cite{Cooperativ_Strategies}, \cite{Gaussian_Parallel}). In
addition, in \cite{Efficient_Protocol} and \cite{User_Cooperation}
the authors proposed cooperation techniques that are based on
extending the relay channel to multiple sources, with information to
transmit, that also serve as relays for each other. The combination
of relaying and cooperation is also an efficient feasible solution
to overcome the channel severity. It is worth mentioning that all of
the models in these references fall within the broader class of
channels with generalized feedback \cite{Feedback_1982},
\cite{Generalized_Feedback}.

 The understanding
of the benefits of MIMO (multiple-input-multiple-output) systems in
wireless channels make the community realize that multiple relays
can emulate the strategies designed for MIMO systems and offer
significant network performance enhancements in terms of various
metrics. These enhancement include increased capacity, extended UAVs
mission range, and improved reliability (by decreasing the outage
probability). This interest has therefore motivated researchers to
analyze the statistics of cooperative relay-based fading channels.
The end-to-end performance in a two-hop system was analyzed in
\cite{End_to_End} over Rayleigh fading channels. However, the
analysis did not consider more than one identical type of fading,
 nor did it consider multiple sources or multiple carrier system.
In this paper, we focus on analyzing the statistics of the aggregate
rates from multiple sources with sub-carriers assignment. The
sub-carriers are assumed to suffer from independent but not
necessarily the same type of fading. More precisely, the statistics
of achievable bit rate of cooperative multi-carrier relay-based
system over generalized fading channels are considered.

The outage probability that gives information about the reliability
of the system is derived in closed form as cascaded weighted sums of
Q-functions. The probability density function (pdf) of achievable
bit rate over various fading channels, including Rayleigh, Weibull,
and Nakagami-\emph{m} fading channels, are expressed as a finite sum
of Gaussian pdfs. The decomposition can be performed using a
well-known procedure called expectation maximization algorithm
\cite{Finite_Mixture_Models}, \cite{Computional_Statistic_Matlab}.
To the best of our knowledge, studying the outage probability and
the average achievable bit rate for such cooperative multi-carrier
framework has not been reported in the literature. The remainder of
this paper is organized as follows. Section II introduces the system
and channel model under consideration. Finite mixture with
expectation maximization algorithm is described in section III.
Then, the problem formulation and expressions for the outage
probability and the average achievable bit rate are given in section
IV. The results are validated using Monte Carlo simulation in
Section V. Finally, the paper is concluded in Section VI. The
notations throughout the paper are summarized in Table
\ref{NotationTable}.
\section{System and Channel Model }
\begin{figure}[p]
\centering
\includegraphics[width=5in]{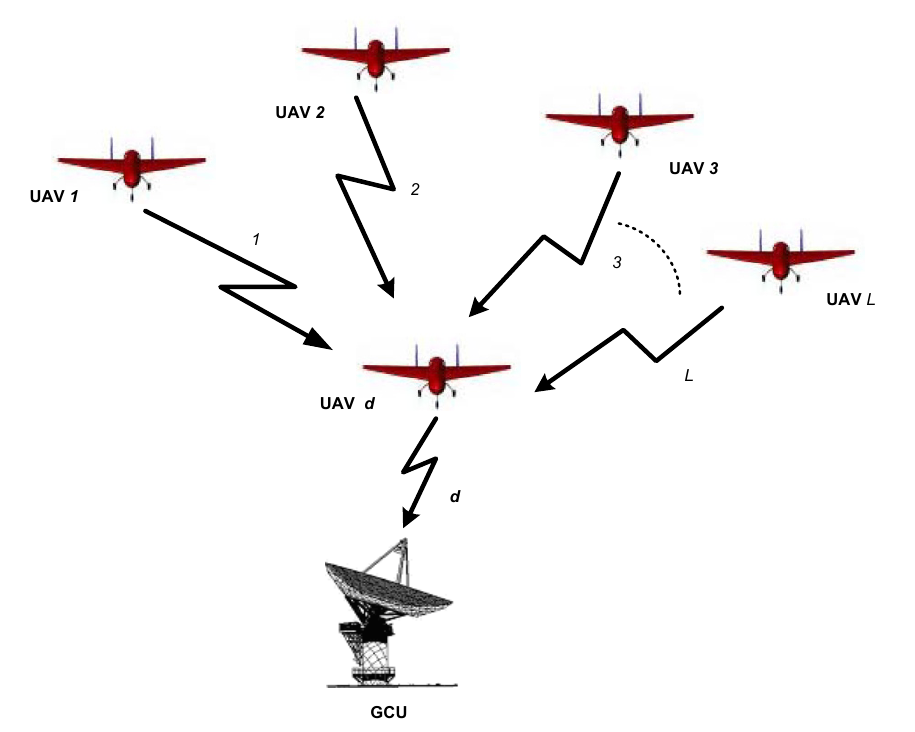}
\caption{Downlink Communication in a cooperative multi-carrier
relay-based UAVs network.} \label{Fig1}
\end{figure}

Consider a group of  cooperative relay-based UAVs located in one
spatial layer as depicted in Fig. \ref{Fig1}.
 In downlink communication, the UAVs are sending their data to the ground control unit (GCU).
 In this topology, each relayed UAV (e.g., UAV $1$ in Fig. \ref{Fig1})
 transmits its data to the relay UAV (UAV $k$ in Fig. \ref{Fig1}).
 The relay UAV in turn carries over the gathered information to the  GCU.
 This topology has the advantage of consuming less power than when data is transmitted directly without relaying.
 However, the relay UAV $d$ consumes more power as compared to
 the relayed ones. Therefore, with good resource management the role of the relay can be exchanged
 between the cooperative UAVs  to achieve fair power consumption. The UAV topology scheduling is beyond
 the scope of this paper. Each UAV requires one uplink channel through which the UAV flight control
data and control information of the on-board sensor payload are
transmitted. In the downlink direction, two channels are utilized.
One provides the position of the UAV, its flight path and navigation
data as well as the internal state of the UAV and the sensor
payload. The other is responsible for providing real time
transmission of the captured video data.

In this paper, the analysis of outage probability and the average
achievable bit rate are considered in the downlink communication as
shown in Fig. \ref{Fig1} with $L$ different links. The total
available bandwidth ($BW$) is divided into $C$ sub-channels (or
equivalently $C$ sub-carriers), where $C>L$. The bit rate over an
$l_{th}$ link depends on the number of sub-carriers ($|C_l|\leq C$)
assigned to that link and is denoted by $R_{l}$, where $l=1,2,...,L$
and $|C_l|$  is the cardinality of subset $C_l$. Furthermore, each
UAV needs to transmit its data with a certain minimum bit rate,
$R_{l,min},~l=1,2,...,L$, that depends on the application the UAV is
used for. It is assumed that the relay-GCU (say the $d_{th}$) link
carries all the data collected from other UAVs besides its own data.
The sub-carriers  $\{c=1,2,...,C\}$ are assigned to the links
$\{l=1,2,...,L\}$, where each sub-carrier is assumed to suffer from
independent but not necessarily identical fading (i.e., the fading
amplitudes of signal transmitted over the sub-carriers are assumed
to be independent random variables).

Three types of fading channel modeling, Rayleigh, Nakagami-\emph{m},
and Weibull fading are considered in this paper. For more
information about these types of fading the reader is refered to
\cite{Fading_Alouni}. The choice of such types of fading can be
considered in cooperative UAVs applications, where for example, in
the low altitude crowded  areas applications, the link may suffer
from Rayleigh fading. On the other hand, most UAV applications are
operated in open space (e.g., at high altitudes) where Nakagami-$m$
and Weibull fading with high fading parameters are suitable fading
models. In this paper, the outage probability is derived using
Gaussian-Finite-Mixture representation, that helps us to represent
the bit rate pdfs as cascaded weighted sums of Gaussian pdfs with
suitable parameters (i.e., means and variances) and weighting
coefficients. In the next section we provide brief description of
Finite-Mixture with expectation maximization algorithm.

\section{Finite Mixture with Expectation Maximization Algorithm}
Finite mixture is a technique for estimating the pdf of a random
variable using given statistical samples. In finite mixture
estimation, it is assumed that a given pdf $f_{\tilde{x}}(x)$ can be
estimated as a weighted sum of a $g$ number of other pdfs. The
parameters of those pdfs can be estimated using $n$ samples (where
 $g<<n$). For the univariate case, the estimated pdf of a random
variable $\tilde{x}$ is given as \cite{Computional_Statistic_Matlab}
\begin{equation}\label{Equation_1}
  f_{\tilde{x}}(x)\simeq \sum_{i=1}^{g}w_i\Phi_i(x;\mathbf{\mathbf{\widehat{\theta}}_\emph{i}}),
\end{equation}
where $w_i$ represents the weighting (mixing) coefficient of the
$i\emph{th}$ term and $
\Phi_i(x;\mathbf{\mathbf{\widehat{\theta}}_\emph{i}})$ denotes the
pdf with parameters represented by the vector
$\mathbf{\widehat{\theta}_\emph{i}}$. An important constraint on
this estimation is to have $w_i>0$ and $\sum_{i=1}^{g}w_i=1$ to
satisfy the unity integral of $f_{\tilde{x}}(x)$. The problem of
estimating the parameters and the weighting coefficients via
different techniques has been considered extensively in literature
(\cite{Computional_Statistic_Matlab} and references therein). One
commonly used estimation technique is the expectation maximization
algorithm \cite{Finite_Mixture_Models}. In order to use expectation
maximization algorithm, we must determine the number of components,
$g$, in the finite mixture model for required accuracy, and initial
estimates for the parameters and the weighting coefficients. Once we
have initial estimates we update the parameters using iterative
updating equations. These equations are derived from the following
equality \cite{Finite_Mixture_Models}:

\begin{equation}\label{Equation_2}
\sum_{j=1}^{n}\tau_{ij}^{(k)}\frac{\partial}{\partial{
z_i}}\sum_{i=1}^g\log\left[
w_i\Phi_i(y_j;\mathbf{\widehat{\theta}_\emph{i}})\right]=0,
\end{equation}
where $z_i$ can be either $w_i$ or
$\mathbf{\widehat{\theta}_\emph{i}}$ depending on either we are
estimating $w_i$ or $\mathbf{\widehat{\theta}_\emph{i}}$,
$\tau_{ij}^{(k)}$ is the $k_{th}$ estimated posteriori probability
that the sample point $y_j$ belongs to the $i_{th}$ weighted pdf,
which can be calculated using \cite{Finite_Mixture_Models}
\begin{equation}\label{Equation_3}
\tau_{ij}^{(k)}=\frac{w_i^{(k)}\Phi_i(y_j;\mathbf{\widehat{\theta}_\emph{i}}^{(k)})}{f_{\tilde{y}}(y_j)}~,
~~ i=1,2, \ldots,g~~,~~ j=1,2, \ldots, n,
\end{equation}
where $g$ is the number of weighted pdfs and $n$ is the total number
of sample points. Throughout the paper, the weighted normal
(Gaussian) pdfs are considered. Then, the estimated pdf is given as
\begin{equation}\label{Equation_4}
f_{\tilde{y}}(y)\simeq\sum_{i=1}^{g}\frac{w_i}{\sqrt{2\pi}\sigma_i}\exp\left(
-\frac{(y-\mu_i)^2}{2\sigma_i^2}\right),
\end{equation}
where $\mu_i$ and $\sigma_i^2$ represent the  mean and the variance,
respectively, of the \emph{ith} weighted pdf which are the
components of the vector $\mathbf{\widehat{\theta}_\emph{i}}$ in
(\ref{Equation_2}). Given the constraint $\sum_{i=1}^{g}w_i=1$ and
using (\ref{Equation_2}), the updating equation to estimate $w_i$,
can be derived as follows:
\begin{align}\label{Equation_5}
&\sum_{j=1}^{n}\tau_{ij}^{(k)}\frac{\partial}{\partial
w_i}\left[\log\left[ \frac{w_i}{\sqrt{2\pi}\sigma_i}\exp\left(
-\frac{(y_j-\mu_i)^2}{2\sigma_i^2}\right)\right]\right] \notag\\
&+ \frac{\partial}{\partial
w_i}\left[\lambda\left[ \sum_{i=1}^gw_i-1\right] \right]= 0\notag\\
&\Longrightarrow
\sum_{j=1}^{n}\tau_{ij}^{(k)}+w_i\lambda=0\Longrightarrow
 \sum_{j=1}^{n}\sum_{i=1}^n\tau_{ij}^{(k)}+\lambda\sum_{i=1}^nw_i=0\notag \\
&\Longrightarrow\sum_{j=1}^{n}1+\lambda=0\Longrightarrow\lambda=-n\notag\\
&\Longrightarrow
w_i^{(k+1)}=\frac{\sum_{j=1}^{n}\tau_{ij}^{(k)}}{n},
\end{align}
where $\lambda$ is the Lagrange multiplier associated with the
constraint ($\sum_{i=1}^{g}w_i=1$). The updating equation to
estimate $\mu_i$, can be derived as follows:
\begin{align}\label{Equation_6}
&\sum_{j=1}^{n}\tau_{ij}^{(k)}\frac{\partial}{\partial\mathbf{\mu_i}}\log\left[
\frac{w_i}{\sqrt{2\pi}\sigma_i}\exp\left(
-\frac{(y_j-\mu_i)^2}{2\sigma_i^2}\right)\right]= 0\notag\\
&\Longrightarrow\sum_{j=1}^{n}\tau_{ij}^{(k)}\frac{\partial}{\partial\mathbf{\mu_i}}\left[
\frac{(y_j-\mu_i)^2}{2\sigma_i^2} \right]=0\notag \\
&\Longrightarrow\sum_{j=1}^{n}\tau_{ij}^{(k)}y_j=\sum_{j=1}^{n}\tau_{ij}^{(k)}\mu_i\notag \\
&\Longrightarrow\mu_i^{(k+1)}=\frac{\sum_{j=1}^{n}\tau_{ij}^{(k)}y_j}{\sum_{j=1}^{n}\tau_{ij}^{(k)}}.
\end{align}
Furthermore, the updating equation to estimate $\sigma_i^2$, can be
derived as follows:
\begin{align}\label{Equation_7}
&\sum_{j=1}^{n}\tau_{ij}^{(k)}\frac{\partial}{\partial\mathbf{\sigma_i^2}}\log\left[
\frac{w_i}{\sqrt{2\pi}\sigma_i}\exp\left(
-\frac{(y_j-\mu_i)^2}{2\sigma_i^2}\right)\right]= 0\notag\\
&\Longrightarrow\sum_{j=1}^{n}\tau_{ij}^{(k)}\frac{\partial}{\partial\mathbf{\sigma_i^2}}\left[
-\frac{(y_j-\mu_i)^2}{2\sigma_i^2}-\frac{1}{2}\log(\sigma_i^2) \right]=0\notag \\
&\Longrightarrow\sum_{j=1}^{n}\tau_{ij}^{(k)}\left[\frac{(y_j-\mu_i)^2}{\sigma_i^2}-1 \right]=0\notag \\
&\Longrightarrow\sigma_i^{2(k+1)}=\frac{\sum_{j=1}^{n}\tau_{ij}^{(k)}(y_j-\mu_i^{(k)})^2}{\sum_{j=1}^{n}\tau_{ij}^{(k)}}.
\end{align}

\begin{figure}\centering
\centering
\includegraphics[width=5in]{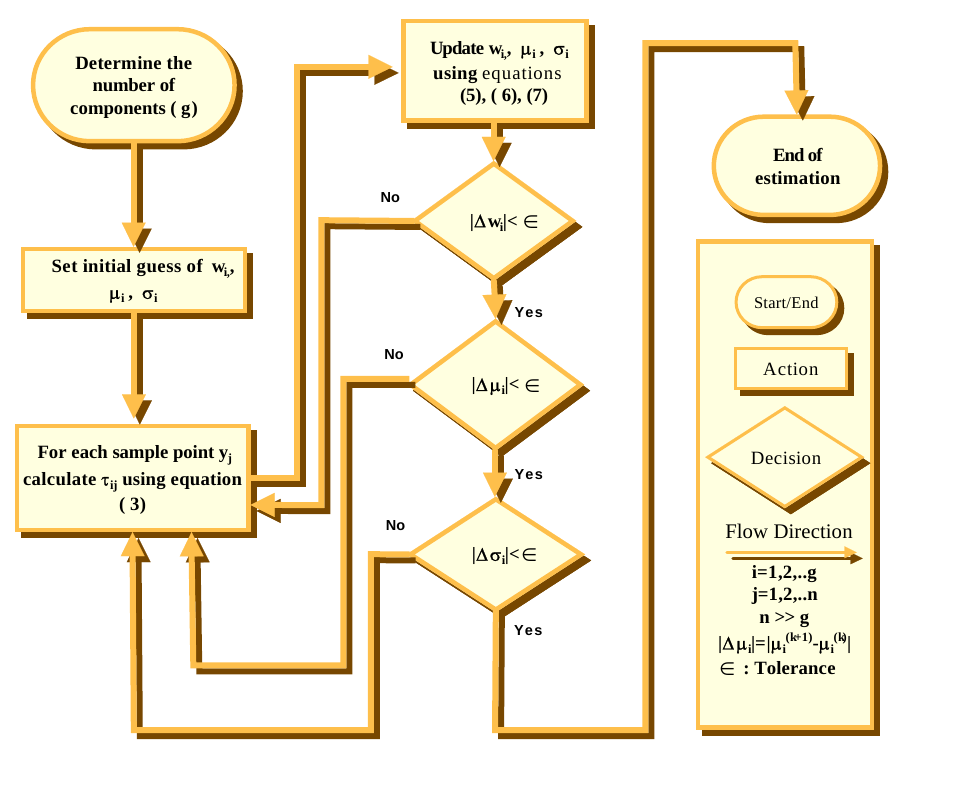}
\caption{Expectation maximization procedural algorithm to estimate
$w_i$, $\mu_i$, and $\sigma_i$ of Gaussian-Finite-Mixture weighted
pdfs.} \label{Flow_Chart_Fig2}
\end{figure}

A procedural description of the expectation maximization algorithm
to estimate $w_i$, $\mu_i$, and $\sigma_i^2$ for $i=1$, $2$, \ldots,
$g$, is shown in Fig. \ref{Flow_Chart_Fig2} (in the figure,
$\epsilon$ is the estimation tolerance). Estimating the pdf of a
positive-valued random variable using weighted Gaussian pdfs could
result in negative part tail for the estimated pdf. Nevertheless,
this tail is negligible and can be truncated with acceptable
accuracy as we will show in the numerical results.

 Typically, the estimation convergence can be implemented by continuing the
iteration until the changes in the estimates at each iteration are
less than some pre-set estimation tolerance $\epsilon$. It is
worthwhile to mention that, the accuracy of estimation depends on
two factors; the number of weighted pdfs, $g$, and the chosen
tolerance $\epsilon$. In addition, the time for convergence
increases with increasing $g$ and with decreasing $\epsilon$. For
more information about the time of convergence and the accuracy of
the expectation maximization algorithm, the reader can refer to
\cite{Finite_Mixture_Models}.


\section{Problem Formulation and Outage Probability analysis}
In the downlink scenario as shown in Fig. \ref{Fig1}, we define the
outage probability of each link as the probability that the link can
not support a minimum required bit rate. In Fig.\ref{Fig1} the
outage probability on the $l_{th}$ and $d_{th}$ links are given by
\begin{equation}\label{Equation_8}
P_{out,l}=P\{R_{l}<R_{l,min}\},\quad l \neq d ,~~l=1,2,\ldots, L,
\end{equation}
\begin{equation}\label{Equation_9}
P_{out,d}=P\left \{R_{d}-\displaystyle \sum_{j=1,j\neq d}^L  R_{j}
\leq R_{d,min} \right
    \}.
\end{equation}
The cooperative relay-based system is declared to be in outage if
one or more of the links are in outage, viz.,
\begin{equation}\label{Equation_10}
   P_{out}=1-\displaystyle \prod_{i=1}^{L}
   \left(1-P_{out,i}\right).
\end{equation}
Adaptive M-ary quadrature amplitude modulation (M-QAM) modulation is
assumed in the analysis, where the spectral efficiency of the
$c_{th}$ sub-carrier over the $l_{th}$ link in $bits/sec/Hz$ is
given as~\cite{adaptive2}
\begin{equation}\label{Equation_11}
   \tilde{r}_{lc}=\log_2(1+\xi\times \tilde{\gamma}_{lc}),
\end{equation}
where $\tilde{\gamma}_{lc}$ is the instantaneous signal-to-noise
ratio (SNR) associated with the $c_{th}$ sub-carrier over the
$l_{th}$ link, $\xi$ is calculated from $\xi=-1.5/\ln(5\times
\mathrm{BER_{target}})$ and $\mathrm{BER_{target}}$ is the required
 bit error rate (BER) which is taken to be $10^{-5}$ in the numerical results throughout the paper.
 $\tilde{\gamma}_{lc}$ can be given mathematically as
 \begin{equation}\label{Equation_12}
  \tilde{\gamma}_{lc}=\frac{E[s^2_{lc}(t)]}{N_o}E[\tilde{a}^2]=\gamma_o E[\tilde{a}^2],
\end{equation}
where $s_{lc}(t)$ is the transmitted signal using the $c_{th}$
sub-carrier over the $l_{th}$ link. $N_o$ is the additive white
gaussian noise (AWGN) one-sided power spectral density (PSD),
$\tilde{a}$ is the fading amplitude of the $c_{th}$ sub-carrier over
the $l_{th}$ link. Approximated pdf of $\tilde{r}_{lc}$ using
$N_{lc}$ Gaussian components can be written as
\begin{equation}\label{Equation_13}
   f_{\tilde{r}_{lc}}(r_{lc})\simeq\sum_{k=1}^{N_{lc}}\frac{w_{lc,k}}{\sqrt{2\pi}\sigma_{lc,k}}\exp \left [ \frac{-(r_{lc}-\mu_{lc,k})^2}{2\sigma_{lc,k}^2} \right ].
\end{equation}
The moment generating function (MGF) of $\tilde{r}_{lc}$ can be
found directly as
\begin{equation}\label{Equation_14}
   \psi_{\tilde{r}_{lc}}(\mathbf{\mathbf{s}})=E[e^{\tilde{r}_{lc}\mathbf{s}}]\simeq\sum_{k=1}^{N_{lc}} w_{lc,k}\exp
   \left [ (\mu_{lc,k}^2) \mathbf{s}  +  (\frac{\sigma_{lc,k}^2}{2}) \mathbf{s}^2 \right ],
\end{equation}
where $E[.]$ is the expectation operation. Let's assume that a set
$C_{l}$ ($l\neq d$) of sub-carriers are assigned to the $l_{th}$
link which is a UAV-relay link, and a set of $C_d$  sub-carriers is
assigned to $d_{th} $ link which is a relay-GCU link, then  the
achievable bit rate over the $l_{th}$ link ($\tilde{R}_{l}$) and the
$d_{th}$ link ($\tilde{R}_{d}$) can be described mathematically by
\begin{equation}\label{Equation_15}
   \tilde{R}_{l}=\sum_{c\in C_{l}}\tilde{r}_{lc}, ~~~~~~~~ l\neq d~,
\end{equation}
\begin{equation}\label{Equation_16}
   \tilde{R}_{d}=\sum_{c\in
C_{d}}\tilde{r}_{dc}-\sum_{l\neq d}\tilde{R}_{l}=\sum_{c\in
C_{d}}\tilde{r}_{dc}-\sum_{l\neq d}\sum _{c \in C_l}\tilde{r}_{lc},
\end{equation}
where $\left\{ C_{d}\bigcup C_{l} :l=1,2,..,L ; l\neq d \right\}$
composes the complete sub-carriers set. Here $d$ is the relay-GCU
link. By assigning a set of sub-carriers (say $C_{l}$) out of $C$
sub-carriers to the $l_{th}$ link and  by assuming independent but
not necessarily identical fading, the MGF of $\tilde{R}_l$ in
(\ref{Equation_15}) can be derived as follows:
\begin{align}\label{Equation_17}
    \Psi_{\tilde{R}_l}(s)&=\prod_{c\in C_l}\Psi_{\tilde{r}_{lc}}(s)\simeq\prod_{c \in C_l} \sum_{k=1}^{N_{lc}}w_{_{lc,k}} \exp
    \left [ \mu_{_{lc,k}}~s+ \frac{\sigma_{_{lc,k}}^2}{2}~s^2 \right ] \notag \\
    &\simeq\sum_{k_{l,1}=1}^{N_{lC_l(1)}}\sum_{k_{l,2}=1}^{N_{lC_l(2)}}....
.
    \sum_{k_{l,n_l}=1}^{N_{lC_l(n_l)}}w_{eq}(k_{l,1},k_{l,2},...,k_{l,n_l})\times \notag \\
    &~~~~~~\exp \left [ \mu_{eq}(k_{l,1},k_{l,2},...,k_{l,n_l})~s+
    \frac{{\sigma^2_{eq}}(k_{l,1},k_{l,2},...,k_{l,n_l})}{2}~s^2 \right ],
\end{align}
where $n_l$ is the number of sub-carriers assigned to $l_{th}$ link
(i.e., $|C_l|$), $C_l(i)$ is the $i_{th}$ sub-carrier assigned to
$l_{th}$ link where $i=1,2,...,n_l$,  $N_{lC_l(i)}$ is the number of
finite mixture Gaussian components (i.e., $g$) used to represent the
pdf of the achievable bit rate, $\tilde{r}_{lC_l(i)}$ associated
with $C_l(i)$ sub-carrier. $w_{eq}(k_{l,1},k_{l,2},...,k_{l,n_l})$,
$\mu_{eq}(k_{l,1},k_{l,2},...,k_{l,n_l})$ and
$\sigma^2_{eq}(k_{l,1},k_{l,2},...,k_{l,n_l})$ are the equivalent
weighting coefficient, mean, and variance, respectively, for the
$l_{th}$ link and can be given as
\begin{equation}\label{Equation_18}
w_{eq}(k_{l,1},k_{l,2},...,k_{l,n_l})=\prod_{m=1}^{n_l}
w_{_{lC_l(m),k_{l,m}}},
\end{equation}
\begin{equation}\label{Equation_19}
\mu_{eq}(k_{l,1},k_{l,2},...,k_{l,n_l})=\sum_{m=1}^{n_l}
\mu_{_{lC_l(m),k_{l,m}}},
\end{equation}
\begin{equation}\label{Equation_20}
\sigma^2_{eq}(k_{l,1},k_{l,2},...,k_{l,n_l})=\sum_{m=1}^{n_l}
\sigma^2_{_{lC_l(m),k_{l,m}}}.
\end{equation}

From (\ref{Equation_17}), which represents an $n_l$ cascaded sums of
Gaussian MGF, one can use inverse Laplace transform directly to
derive the pdf of the achievable bit rate over the $l_{th}$ link,
which can be given as
\begin{equation}\label{Equation_21}
f_{\tilde{R}_l}(r_l)\simeq\sum_{k_{l,1}=1}^{N_{lC_l(1)}}\sum_{k_{l,2}=1}^{N_{lC_l(2)}}...\sum_{k_{l,n_l}=1}^{N_{lC_l(n_l)}}
\frac{w_{eq}(k_{l,1},k_{l,2},...,k_{l,n_l})}{\sqrt{2\pi
}\sigma_{eq}(k_{l,1},k_{l,2},...,k_{l,n_l})}\exp \left [
-\frac{(r_l-\mu_{eq}(k_{l,1} ,k_{l,2},...,k_{l,n_l}))^2}{2
\sigma^2_{eq}(k_{l,1},k_{l,2},...,k_{l,n_l})} \right ].
\end{equation}
The average achievable bit rate ($E[\tilde{R}_l]$) can be derived by
averaging $\tilde{R}_l$ over the pdf in (\ref{Equation_21}) as
follows:
 \begin{align}\label{Equation_22}
E[\tilde{R}_l]&=\int_0^\infty r_l f_{\tilde{R}_l}(r_l) dr_l \\
\notag &\simeq\sum_{k_{l,1}=1}^{N_{lC_l(1)}}
\sum_{k_{l,2}=1}^{N_{lC_l(2)}}...\sum_{k_{l,n_l}=1}^{N_{lC_l(n_l)}}w_{eq}
(k_{l,1},k_{l,2},...,k_{l,n_l})\mu_{eq}(k_{l,1},k_{l,2},...,k_{l,n_l}).
\end{align}

It is noticed in (\ref{Equation_22}) that the integration limits
take only positive values while the Gaussian pdfs are truncated for
$r_l<0$, which is acceptable truncation since the tails of Gaussian
pdfs in (\ref{Equation_21}) are negligible when $r_l<0$. The outage
probability on the $l_{th}$ link can be derived by substituting
$f_{\tilde{R}_l}(r_l)$ from (\ref{Equation_21}) into
(\ref{Equation_8}) and performing the integration to obtain
\begin{align}\label{Equation_23}
    P_{out,l}(R_{l,min})& \simeq \int_0^{R_{l,min}} f_{\tilde{R}_l}(r_l) dr_l \notag \\
       &\simeq \sum_{k_{l,1}=1}^{N_{lC_l(1)}}\sum_{k_{l,2}=1}^{N_{lC_l(2)}} ...
       \sum_{k_{l,n_l}=1}^{N_{lC_l(n_l)}}w_{eq}(k_{l,1},k_{l,2},...,k_{l,n_l}) \times \notag \\
       &~~~~\left [ 1-Q \left( \frac{R_{l,min}-\mu_{eq}(k_{l,1},k_{l,2},...,k_{l,n_l})}{
       \sigma_{eq}(k_{l,1},k_{l,2},...,k_{l,n_l})}\right)
    - Q \left( \frac{\mu_{eq}(k_{l,1},k_{l,2},...,k_{l,n_l})}{\sigma_{eq}(k_{l,1},k_{l,2},...,k_{l,n_l})}\right)\right ],
\end{align}
where $Q(x)$ is the Q-function defined as
\begin{equation}\label{Equation_24}
Q(x)=\int_{x}^{\infty}\frac{1}{\sqrt{2\pi}}\exp(-\frac{u^2}{2})du,
\end{equation}
and $R_{l,min}$ is the minimum bit rate required to be sent over the
$l_{th}$ link. The MGF of $\tilde{R}_d$
 can be derived by using (\ref{Equation_16}) and (\ref{Equation_13}). From (\ref{Equation_16}) we obtain
\begin{align}\label{Equation_25}
    \Psi_{\tilde{R}_d}(s)&=\prod_{c\in C_d}\Psi_{\tilde{r}_{dc}}(s)     \prod_{l \neq d}\Psi_{\tilde{R}_l}(-s)=
     \prod_{c\in C_d}\Psi_{\tilde{r}_{dc}}(s)\prod_{l \neq d}\prod_{c \in
     C_l}\Psi_{\tilde{r}_{lc}}(-s).
\end{align}
To simplify the analysis, let's define a parameter $\alpha_l=1$ if
$l=d$, and  $\alpha_l=-1$ otherwise ($l \neq d$), then using
(\ref{Equation_14}) and after some manipulations (\ref{Equation_25})
becomes
 \begin{align}\label{Equation_26}
    \Psi_{\tilde{R}_d}(s)&=\prod_{l=1}^{L}\prod_{m=C_l(1)}^{C_l(n_l)}
    \Psi_{\tilde{r_l},m}(\alpha_l s) \notag \\
    &\simeq\prod_{l=1}^{L}\sum_{k_{l,1}=1}^{N}\sum_{k_{l,2}=1}^{N}
    ...\sum_{k_{l,n_l}=1}^{N}w_{eq}(k_{l,1},k_{l,2},...,k_{l,n_l}) \times \notag \\
    &\exp \left[ \mu_{eq}(k_{l,1},k_{l,2},...,k_{l,n_l})~\alpha_l s+ \frac{{\sigma^2_{eq}}
    (k_{l,1},k_{l,2},...,k_{l,n_l})}{2}~s^2 \right ],
\end{align}
where in (\ref{Equation_26}) we denoted to the number of Gaussian
components for all sub-carriers over
 all the links by $N$ for simplicity. If we define the operator $\underline{\overline{\sum}}^\mathbb{N}_{\mathbb{K}_l}
 =\sum_{k_{l,1}=1}^{N}\sum_{k_{l,2}=1}^{N}...\sum_{k_{l,n_l}=1}^{N}$, then, the MGF of the achievable bit rate over
 the $d_{th}$ link can be written as
 \begin{align}\label{Equation_27}
    \Psi_{\tilde{R}_d}(s)&=\underline{\overline{\sum}}^\mathbb{N}_{\mathbb{K}_1}
    \underline{\overline{\sum}}^\mathbb{N}_{\mathbb{K}_2}...\underline{\overline
    {\sum}}^\mathbb{N}_{\mathbb{K}_L}\left(\prod_{l=1}^L w_{eq}(k_{l,1},k_{l,2},...,k_{l,n_l})\right ) \times \notag \\
    &~~~~\exp \left [ s\sum_{l=1}^L\mu_{eq}(k_{l,1},k_{l,2},...,k_{l,n_l})~\alpha_l+ s^2 \frac{{\sum_{l=1}^L
    \sigma^2_{eq}}(k_{l,1},k_{l,2},...,k_{l,n_l})}{2} \right ].
\end{align}
From (\ref{Equation_27}) which represents an $n_l^2$ cascaded sums
of Gaussian MGF, the inverse Laplace transform can be used to derive
the pdf of the achievable bit rate over the $d_{th}$ link, which can
be given as
 \begin{align}\label{Equation_28}
   f_{\tilde{R}_d}(r_d)&\simeq\underline{\overline{\sum}}^\mathbb{N}_{\mathbb{K}_1}
    \underline{\overline{\sum}}^\mathbb{N}_{\mathbb{K}_2}...\underline{\overline
    {\sum}}^\mathbb{N}_{\mathbb{K}_L} \frac{\prod_{l=1}^L w_{eq}(k_{l,1},k_{l,2},...,k_{l,n_l})}
    {\sqrt{2\pi\sum_{l=1}^L\sigma^2_{eq}(k_{l,1},k_{l,2},...,k_{l,n_l})}} \times \notag \\
    &~~~~~~~~~~\exp \left [ -\frac{
    \left(r_d-\sum_{l=1}^L\mu_{eq}(k_{l,1},k_{l,2},...,k_{l,n_l}) \alpha_l\right)^2}{
    2\sum_{l=1}^L\sigma^2_{eq}(k_{l,1},k_{l,2},...,k_{l,n_l})} \right ].
\end{align}
 The average achievable bit rate ($E[\tilde{R}_d]$) over the $d_{th}$ link (relay-GCU link) can be derived
 by averaging $\tilde{R}_d$ over the pdf in ($\ref{Equation_28}$). After little manipulation we can show that
   \begin{align}\label{Equation_29}
   E[\tilde{R}_d]=\int_0^\infty r_d f_{\tilde{R}_d}(r_d) dr_d\simeq&\underline{\overline{\sum}}^\mathbb{N}_{\mathbb{K}_1}
    \underline{\overline{\sum}}^\mathbb{N}_{\mathbb{K}_2}...\underline{\overline
    {\sum}}^\mathbb{N}_{\mathbb{K}_L} \prod_{l=1}^L w_{eq}(k_{l,1},k_{l,2},...,k_{l,n_l})\times \notag \\
     &~~~~~~~~~\sum_{l=1}^L\mu_{eq}(k_{l,1},k_{l,2},...,k_{l,n_l}) \alpha_l.
\end{align}

The outage probability on the $d_{th}$ link can be derived by
substituting (\ref{Equation_28}) into (\ref{Equation_9})  given that
the minimum required transmission bit rate is $R_{d,min}$. After
performing the integration, $P_{out,d}(R_{d,min})$ can be given as
 \begin{align}\label{Equation_30}
   P&_{out,d}(R_{d,min})\simeq\underline{\overline{\sum}}^\mathbb{N}_{\mathbb{K}_1}
    \underline{\overline{\sum}}^\mathbb{N}_{\mathbb{K}_2}...\underline{\overline
    {\sum}}^\mathbb{N}_{\mathbb{K}_L} \left(\prod_{l=1}^L w_{eq}(k_{l,1},k_{l,2},...,k_{l,n_l})\right)\notag \\
    &\times\left [ 1-Q \left( \frac{R_{k,min}-\sum_{l=1}^L\mu_{eq}(k_{l,1},k_{l,2},...,k_{l,n_l}) \alpha_l}
    {\sqrt{\sum_{l=1}^L\sigma^2_{eq}(k_{l,1},k_{l,2},...,k_{l,n_l})}}\right)
     - Q \left( \frac{\sum_{l=1}^L\mu_{eq}(k_{l,1},k_{l,2},...,k_{l,n_l})
     \alpha_l}{\sqrt{\sum_{l=1}^L\sigma^2_{eq}(k_{l,1},k_{l,2},...,k_{l,n_l})}}\right) \right].
    \end{align}

 It is noticed that the derived outage probability expressions are given in terms of cascaded
weighted sums of well-known Q-function as shown in
(\ref{Equation_23}) and (\ref{Equation_30}). These expressions
provide a generalized tool to study the reliability of cooperative
multi-carrier relay-based UAVs network over variety types of
wireless fading channels. In the following section, numerical
results and Monte Carlo simulation are used to validate theses
expressions.

\section{Numerical Results}
 Cooperative multi-carrier relay-based UAVs network with three links $1$, $2$, and $3$ is
 considered as shown if Fig.~\ref{Fig1}, where the $1^{st}$ link represents the relay-GCU link. The system is
  assumed to operate over a $BW$ of 80 MHz  which is divided into 10 sub-channels
  (8 MHz each) in the 2.4 GHz industrial, scientific, and medical (ISM) band. The number of sub-channels
  (or equivalently sub-carriers) assigned to links $1$, $2$,  and $3$ are $C_1=6$, $C_2=2$,  and $C_3=2$,
  respectively. Five fading scenarios that include three types of fading; Rayleigh, Nakagami-$m$, and
  Weibull fading, are considered in the simulation. The five fading scenarios are summarized in Table
  \ref{FadingScenarios}. In our three-links example, the sub-carriers allocation and
  their associated fading scenarios are summarized in Table \ref{ChannelAllocation}. The allocated sub-carriers are
   assumed to have different fading scenarios to demonstrate the feasibility and accuracy of the approximated analytical
   expressions as compared with Monte Carlo simulation where we performed $10^5 $ simulation runs. Four Gaussian-Finite-Mixture components (i.e., $g=4$) were shown to be enough to provide accuracy up to
    $\epsilon=10^{-3}$ to  approximate the pdf of the achievable bit rate in $bits/sec/Hz$ for the  five fading
  scenarios described in Table \ref{FadingScenarios}. The results are categorized into three groups, the first group
  (Fig. 3, Fig. 4, Fig. 5, Fig. 6, and Fig. 7) are the results for one sub-channel per link in five different fading scenarios.
  The second group (Fig. 8, Fig. 9, and Fig. 10) demonstrates the accuracy of approximated expressions by increasing the number of sub-channels per link.
  The third group of results (Fig. 11 and Fig. 12) present the approximated expressions (i.e., the achievable
  rate and the outage probability) in a three links example.

\begin{figure}[p] \centering
\includegraphics[width=5in]{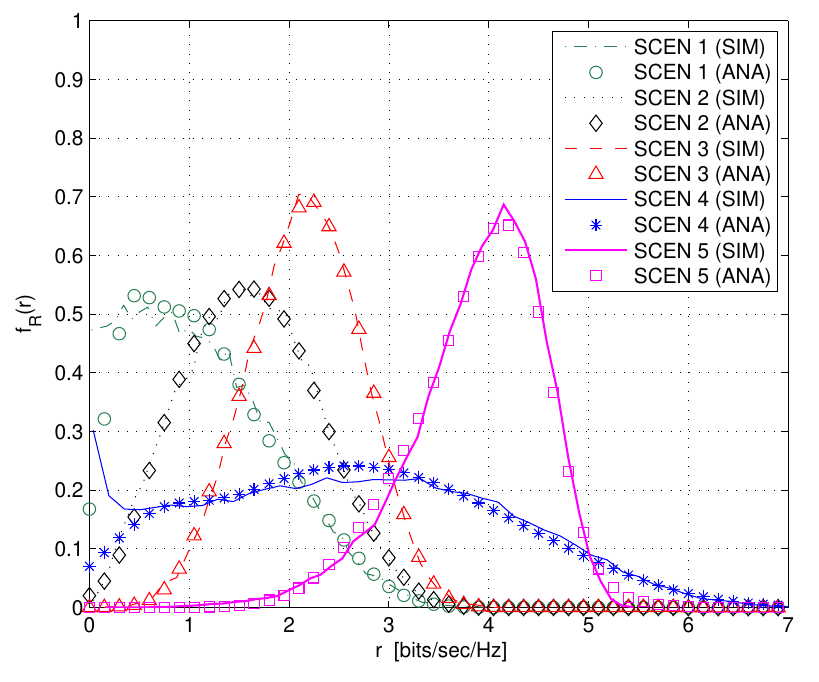}
\caption{Comparison between the analytical results (ANA) of the achievable rate pdf ($f_{\tilde{R}}(r)$) and
the Monte Carlo simulation (SIM) for various fading scenarios as given in Table \ref{FadingScenarios}.}
\label{Fig1_Rate_PDF_5Scen}
\end{figure}

\begin{figure}[p] \centering
\centering
\includegraphics[width=5in]{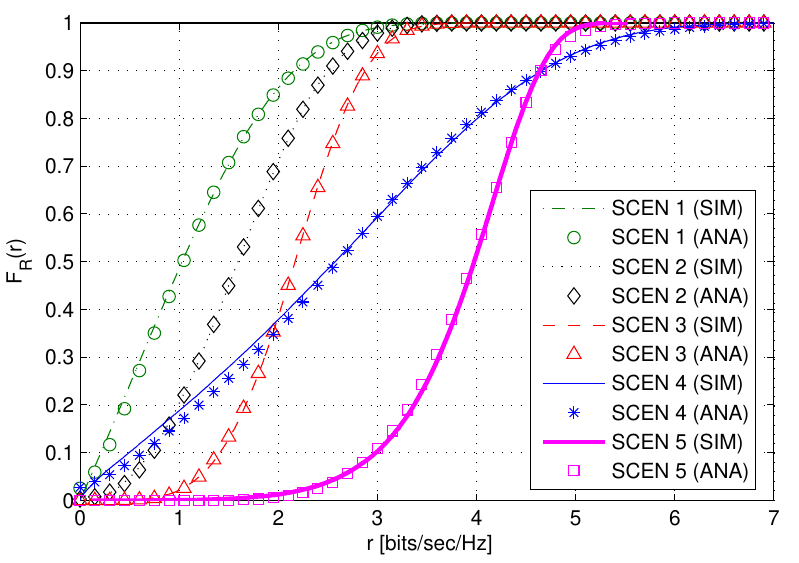}
\caption{Comparison between the analytical results (ANA) of the achievable
 rate CDF ($F_{\tilde{R}}(r)$) and the Monte Carlo simulation (SIM) for various fading scenarios as given
 in Table \ref{FadingScenarios}.}
\label{Fig2_Rate_CDF_5Scen}
\end{figure}

  Fig. \ref{Fig1_Rate_PDF_5Scen} gives a comparison between
   the analytical (approximated) results for the pdf of the achievable bit rate, $f_{\tilde{R}}(r)$, and the
   Monte Carlo simulation for various fading scenarios as given in Table \ref{FadingScenarios}.
  It is obvious from Fig. \ref{Fig1_Rate_PDF_5Scen} that the pdf of the achievable bit rate of fading
   scenarios $2,3,5$ can be approximated to a high degree of accuracy using four components ($g=4$),
   where we still need larger number of components to estimate, accurately, the other two fading scenarios;
    Rayleigh fading (SCEN $1$) and  Weibull fading with $\beta=1.5$ (SCEN $2$). The picture becomes more
    clear by investigating the CDF and the MGF of the achievable bit rate. Figs. \ref{Fig2_Rate_CDF_5Scen}
    and \ref{Fig3_Rate_MGF_5Scen}  compare the analytical (approximated) results of the CDF
     and MGF  of the achievable bit rate, respectively, and the Monte Carlo simulation for various fading
     scenarios as given in Table \ref{FadingScenarios}. Fig. 5 and  Fig. 6 to show the accuracy of approximated pdf and CDF expression
      as function of g (i.e., the number of Finite Mixture Components). Although, the approximation is not highly accurate for
     severe fading scenarios (i.e., SCEN 1), but we still can achieve very accurate approximation (i.e., $|error|< 3\%$ ) by using $g>=4$
     as shown in Fig. 5.

      Fig. \ref{Fig4_Rate_Vs_Go_5Scen_2_4_SCH} Compares the analytical results  of the achievable
      bit rate in (\ref{Equation_22}) and the Monte Carlo simulation for various fading scenarios using two and four 8 MHz sub-channels.
  It is well noticed that on average the analytical results coincide to high degree of accuracy ($|$Error$|$$<3$$\%$)
  with the simulation. The analytical results of the outage probability in (\ref{Equation_23})
  using two and four 8 MHz sub-channels are compared with simulation for various fading scenarios and are given in
  Fig. \ref{Fig5_Out_1Link_2SCH} and Fig. \ref{Fig6_Out_1Link_4SCH}, respectively. Here, the general conclusion
  is that the outage probability expression in terms of a cascaded weighted sums of Q-functions in ($\ref{Equation_23}$)
  can be used with high degree of accuracy to analyze the outage probability of a cooperative multi-carrier
  relay-based system for various fading scenarios. For the three-links example that is shown in
   Fig. \ref{Fig1} and described by Table \ref{ChannelAllocation}, the average
   achievable bit rate over the three links as in (\ref{Equation_22}) and
   (\ref{Equation_29}) are calculated and compared with simulation
   as shown in Fig. \ref{Fig7_Rate_Vs_Go_3Links_10_SCH}. As last investigation, the outage probability in
   (\ref{Equation_23}) and (\ref{Equation_30}) are plotted in Fig. \ref{Fig8_Out_3Links_10_SCH} and are compared
    with simulation. It is clear from Fig. \ref{Fig7_Rate_Vs_Go_3Links_10_SCH} and \ref{Fig8_Out_3Links_10_SCH}
    to a high degree of accuracy the simulations coincide with the analytical results.

\section{Conclusion}

In this paper, the reliability analysis in terms of outage
probability of cooperative  multi-carrier relay-based UAVs network
over generalized fading channels is considered. The average
achievable bit rate for each link is also derived in closed form.
General analytical expressions in terms of cascaded weighted sums of
well-known Q-function for the outage probability of both UAV-relay
and  relay-GCU links are derived. The expressions are derived for
the case of independent but not necessary identically distributed
fading channels. These expressions provide a generalized tool to
study the reliability of cooperative multi-carrier relay-based UAVs
network over variety of fading scenarios. In our numerical
demonstrations, we used Weibull, Nakagami-\emph{m}, and Rayleigh
fading channels as examples. The expressions were validated using
Monte Carlo simulation.


\begin{table}[h]
\caption{Table of Notations} \centering
\begin{tabular}{||l|l||}
  \hline
  \hline
  Notation & Description \\
  \hline
   \hline
 $C$ & Number of sub-channels  \\
  \hline
 $L$ & Number of links \\
 \hline
 $R_{l,min}$ & Minimum bit rate over the $l_{th}$ link \\
  \hline
  $\tilde{R}_{l}$ & Bit rate over the $l_{th}$ link\\
  \hline

  $\emph{m}$ & Nakagami-\emph{m} fading parameter \\
  \hline
 $\beta$ & Weibull fading parameter \\
 \hline

 $g$ & Number of Finite Mixture components\\
  \hline

  $n$ & Number of samples \\
  \hline

   $w_i$ & Weighting coefficient of the $i_{th}$ term  \\
  \hline

$\Phi_i(x;\mathbf{\mathbf{\widehat{\theta}}_\emph{i}})$ &The $i_{th}$ pdf with parameters vector $\mathbf{\mathbf{\widehat{\theta}}_\emph{i}}$ \\
 \hline

   $\mu_i$ & The mean of the $i_{th}$ weighted pdf \\
  \hline

$\sigma_i^2$ & The variance of the $i_{th}$ weighted pdf \\
 \hline

   $\epsilon$ & Estimation tolerance \\
  \hline

$\tilde{r}_{lc}$ & Achievable data rate associated with the $c_{th}$
sub-carrier over the $l_{th}$ link \\
 \hline

 $\lambda$ & Lagrange multiplier \\
  \hline

$P_{out,l}$ & Outage probability of the $i_{th}$ link \\
\hline

$\tilde{\gamma}_{lc}$ & The SNR associated with the $c_{th}$
sub-carrier over the $l_{th}$ link \\
\hline $s_{lc}(t)$ & The transmitted signal using the $c_{th}$ sub-carrier over the $l_{th}$ link \\
\hline

$\Psi_{\tilde{R}_l}(s)$ & The MGF of $\tilde{R}_l$ \\
\hline

$f_{\tilde{R}_d}(r_d)$ & The pdf of the achievable bit rate over the $d_{th}$ link \\
\hline

$E[.]$ & The expectation operation \\
\hline

$F_{\tilde{R}_d}(r_d)$ & The CDF of the achievable bit rate over the $d_{th}$ link \\
\hline
$\underline{\overline{\sum}}^\mathbb{N}_{\mathbb{K}_l}$ & $\sum_{k_{l,1}=1}^{N}\sum_{k_{l,2}=1}^{N}...\sum_{k_{l,n_l}=1}^{N}$ where $\mathbb{K}_l=[k_{l,1}, k_{l,2}.., k_{l,n_l}]$ \\

\hline
\hline
\end{tabular}\label{NotationTable}
\end{table}

\begin{table}[h]
\caption{Fading Scenarios }
\centering
\begin{tabular}{|l|l|l|l|}
  \hline
  \hline
  Scenario & Fading &Parameter&$\gamma_o$ [dB] \\
  \hline
   \hline
  SCEN 1 & Rayleigh &$E[\tilde{a}^2]=1$&10 \\
  \hline
  SCEN 2 & Nakagami-$m$ &$E[\tilde{a}^2=1]=1$, $m=2$&12 \\
  \hline
  SCEN 3 & Nakagami-$m$ &$E[\tilde{a}^2=1]$, $m=4$& 14 \\
   \hline
  SCEN 4  & Weibull &$E[\tilde{a}^2]=1$, $\beta=1.5$&18 \\
  \hline
  SCEN 5  & Weibull &$E[\tilde{a}^2]=1$, $\beta=5$&20 \\
  \hline

\end{tabular}\label{FadingScenarios}
\end{table}

\begin{table}[h]
\caption{Sub-carriers allocation and fading scenarios }
\centering
\begin{tabular}{|l|l|l|}
  \hline
  \hline
  Link & Sub-channel No.& Fading Scenario  \\
  \hline \hline
  1 &1,2,3,4,5,6& 1,2,3,4,5,5\\
  \hline
  2 &7,8&2,2\\
  \hline
  3 &9,10&3,4\\
  \hline
\end{tabular}\label{ChannelAllocation}
\end{table}


\begin{figure}[p] \centering
\centering
\includegraphics[width=5in]{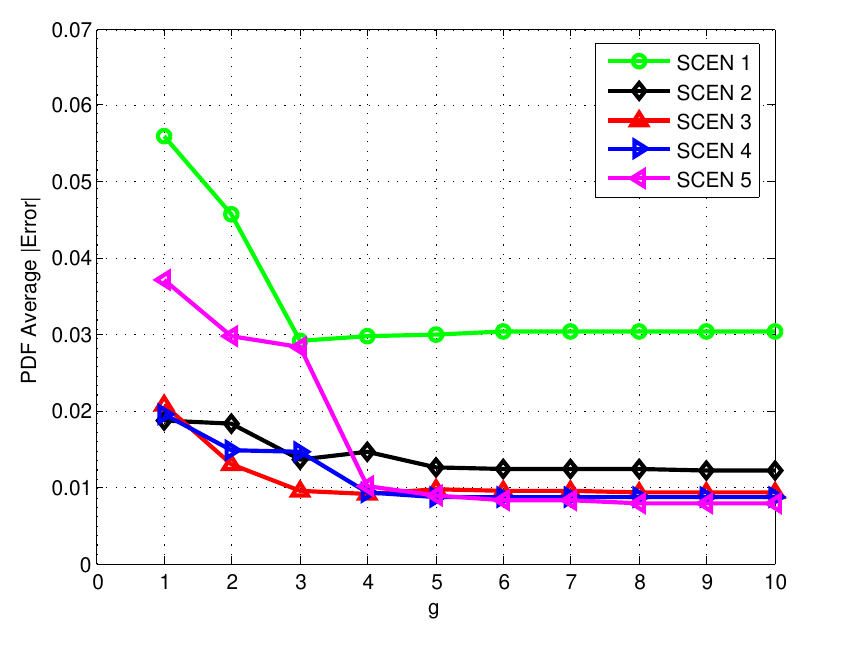}
\caption{Average absolute error in the achievable rate pdf
expression ($f_{\tilde{R}}(r)$) as compared to the Monte Carlo
simulation for different number of components (\emph{g}).}
\label{Fig3_PDF_Error}
\end{figure}

\begin{figure}[p] \centering
\centering
\includegraphics[width=5in]{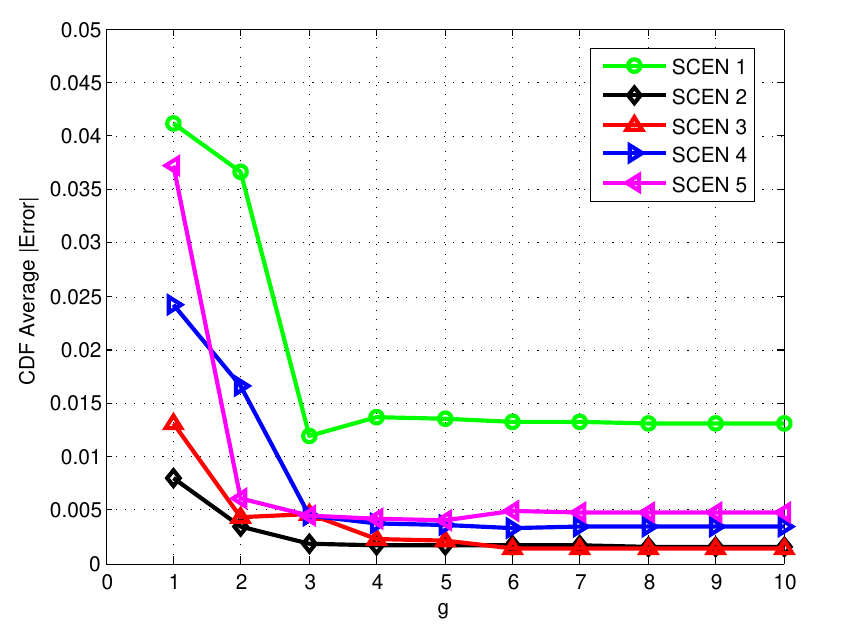}
\caption{Average absolute error in the achievable CDF rate
expression ($F_{\tilde{R}}(r)$) as compared to the Monte Carlo
simulation for different number of components (\emph{g}).}
\label{Fig4_PDF_Error}
\end{figure}

\begin{figure}[p]  \centering
\centering
\includegraphics[width=5in]{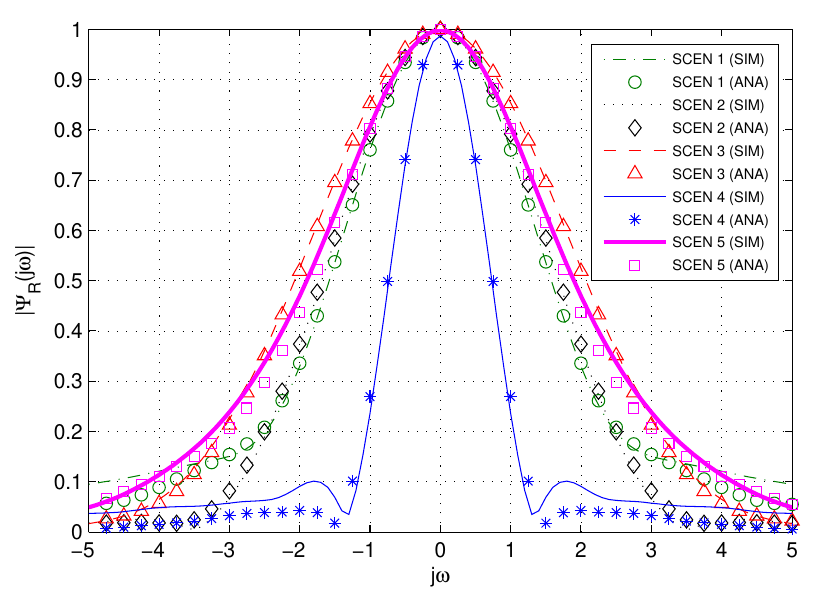}
\caption{Comparison between the analytical results (ANA) of the MGF
of the achievable rate  $|\Psi_{\tilde{R}}(jw)|$ and the Monte Carlo simulation (SIM) for various fading scenarios
as given in Table \ref{FadingScenarios}. }
\label{Fig3_Rate_MGF_5Scen}
\end{figure}

\begin{figure}[p] \centering
\centering
\includegraphics[width=5in]{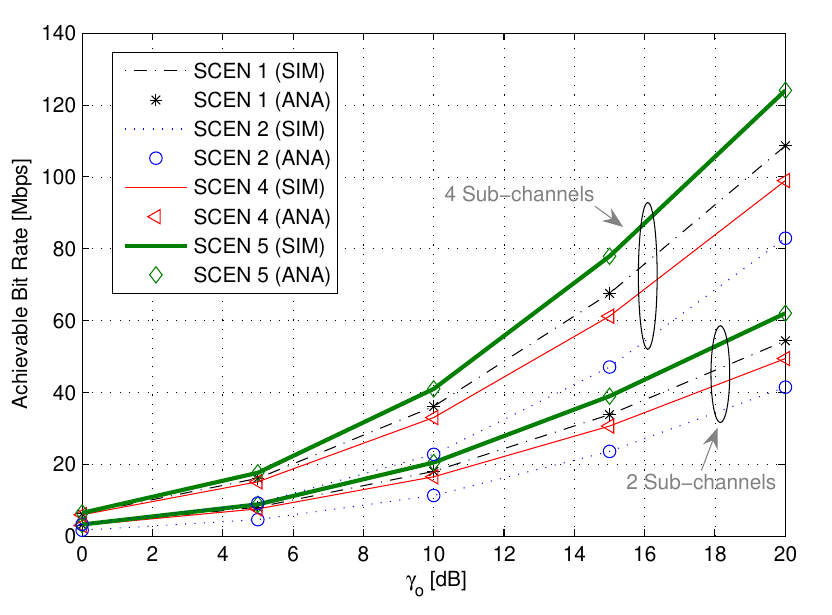}
\caption{Comparison between the analytical results (ANA) of the achievable rate ($E[\tilde{R}]$) and the Monte
Carlo simulation (SIM) for various fading scenarios using two  and four 8 MHz sub-channels.}
\label{Fig4_Rate_Vs_Go_5Scen_2_4_SCH}
\end{figure}

\begin{figure}[p] \centering
\centering
\includegraphics[width=5in]{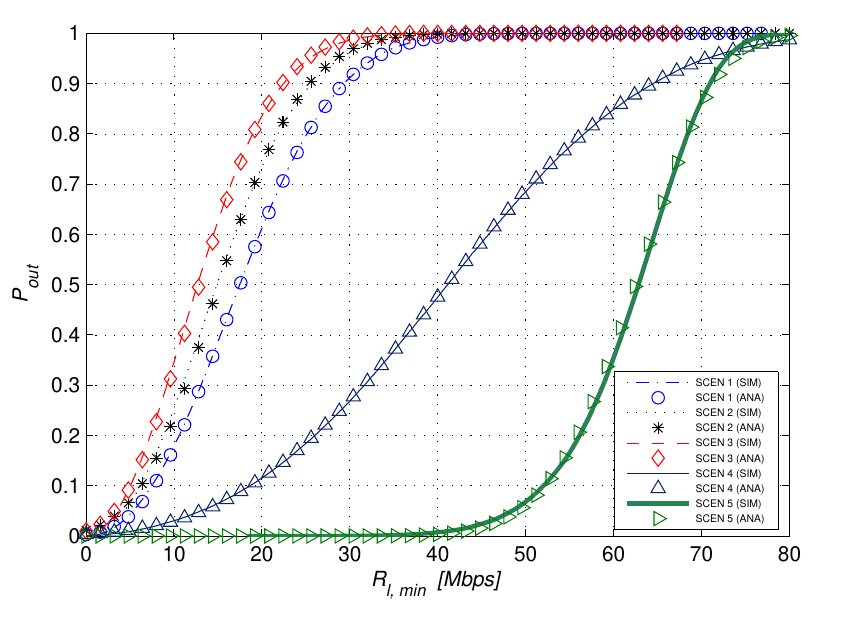}
\caption{Comparison between the analytical results (ANA) for the outage probability
$P_{out}$ in (\ref{Equation_23})
and the Monte Carlo simulation (SIM) for various fading scenarios using two 8 MHz sub-channels.}
\label{Fig5_Out_1Link_2SCH}
\end{figure}

\begin{figure}[p] \centering
\centering
\includegraphics[width=5in]{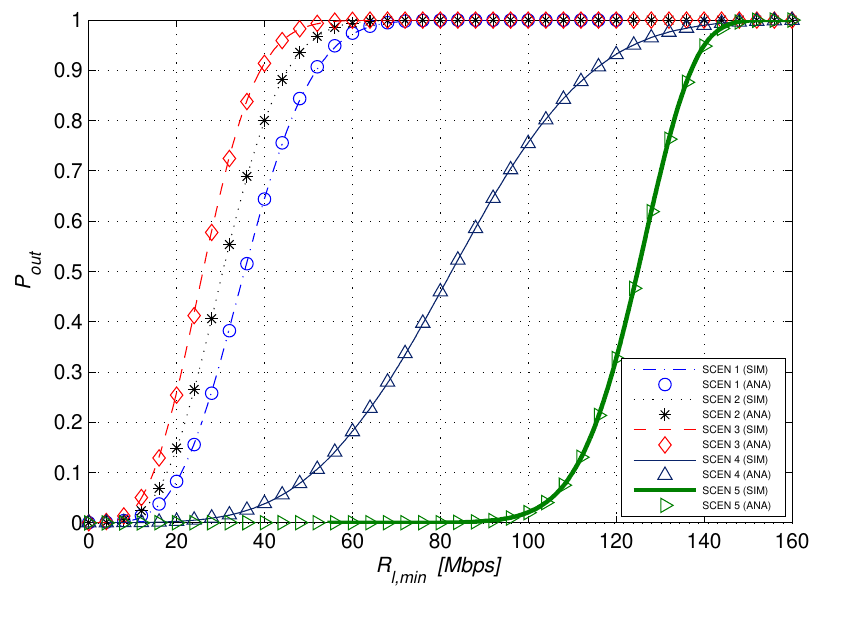}
\caption{Comparison between the analytical results  (ANA) for the outage probability $P_{out}$ in (\ref{Equation_23})
and the Monte Carlo simulation (SIM) for various fading scenarios using four 8 MHz sub-channels.}
\label{Fig6_Out_1Link_4SCH}
\end{figure}

\begin{figure}[p] \centering
\centering
\includegraphics[width=5in]{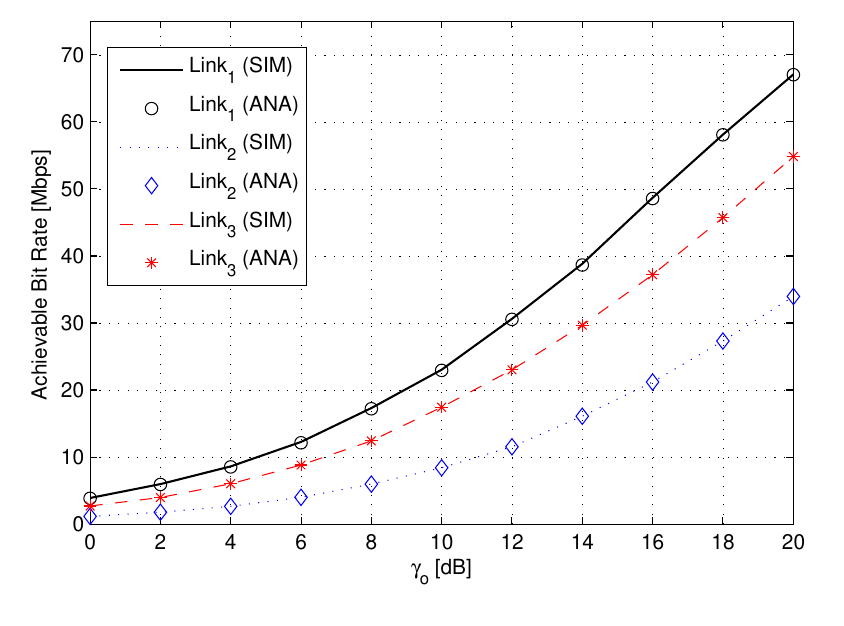}
\caption{Comparison between the analytical results (ANA) in
(\ref{Equation_22}) and (\ref{Equation_29}) for the achievable rate
and the Monte Carlo simulation for the three links example given in
Fig. \ref{Fig1} and described in Table \ref{ChannelAllocation}.}
\label{Fig7_Rate_Vs_Go_3Links_10_SCH}
\end{figure}

\begin{figure}[p] \centering
\centering
\includegraphics[width=5in]{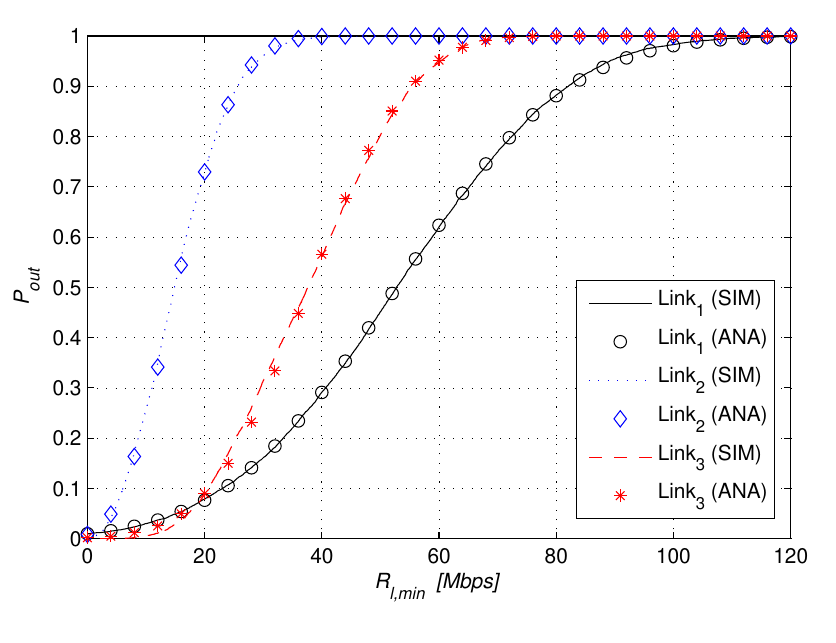}
\caption{Comparison between the analytical results(ANA) in
(\ref{Equation_23}) and (\ref{Equation_30}) for the outage
probability and the Monte Carlo simulation (SIM) for the three links
example given in Fig. \ref{Fig1} and described in Table
\ref{ChannelAllocation}.} \label{Fig8_Out_3Links_10_SCH}
\end{figure}


\begin{thebibliography}{1}
\bibitem{UAV_Book_1}
P. Castillo, R. Lozano and E. DZUL, \emph{Modelling and Control of
Mini-Flying Machines}.
 London: Springer, 2005.

\bibitem{UAV_Book_2}
L. R. Newcome, \emph{Unmanned Aviation: A Brief History of Unmanned
Aerial Vehcles}. Verginia: American Institute of Aeronautics \& Ast,
2004.

\bibitem{UAV_Book_3}
K.P. Valavanis, \emph{Advances in Unmanned Aerial Vehicles},
Springer, Sep 2007.

\bibitem{combinedDSSSFHSS}
M. Edrich and R. Schmalenberger, ``Combined DSSS/FHSS approach to
interference rejection and navigation support in UAV communications
and control," in \textit{Proc. IEEE 7th Int. Symp. on Spread
Spectrum Techn. and App.}, Prague, Czech Republic, vol. 3, 2002, pp.
687--691.
\bibitem{Capacity_Theorems}
T. Cover and A. Gamal, ``Capacity theorems for the relay channel,"
\emph{IEEE Trans. Inform. Theory}, vol. 25, no. 5, pp. 572-584,
Sep. 1979.

\bibitem{Three_Terminals_71}
E.C. van der Meulen, ``Three-terminal communication channels,"
\emph{Adv. Appl. Prob.}, vol. 3, pp. 120-154, 1971.
\bibitem{Cooperativ_Strategies}
G. Kramer, M. Gastpar, and P. Gupta, ``Cooperative strategies and
capacity theorems for relay networks," \emph{IEEE Trans. Inform.
Theory}, vol. 51, no. 9, pp. 3037-3063, Sep. 2005.
\bibitem{Gaussian_Parallel}
B. Schein and R.G. Gallager, ``The Gaussian parallel relay network,"
\emph{in Proc. IEEE Int. Symp. Inform. Theory} (ISIT), Sorrento,
Italy, June 2000
\bibitem{Efficient_Protocol}
J. N. Laneman, D. C. Tse and G. W. Wornell, ``Cooperative Diversity
in Wireless Networks: Efficient Protocols and Outage Behavior,"
\emph{IEEE Trans. Inform. Theory}, vol. 50, no. 12, pp. 3062-3080,
Dec. 2004.
\bibitem{User_Cooperation}
A. Sendonaris, E. Erkip, and B. Aazhang, ``Increasing uplink
capacity via user cooperation diversity," \emph{in Proc. IEEE Int.
Symp. Inform. Theory}, Cambridge, MA, Aug. 1998
\bibitem{Feedback_1982}
A.B. Carleial, ``Multiple-access channels with different generalized
feedback signals," \emph{IEEE Trans. Inform. Theory}, vol. 28, no.
6, pp. 841-850, Nov. 1982.
\bibitem{Generalized_Feedback}
R.C. King, ``Multiple access channels with generalized feedback,"
Ph.D. dissertation, Stanford University, Palo Alto, CA, Mar. 1978.

\bibitem{End_to_End}
M.O. Hasna, M. Alouini, ``End-to-end performance of transmission
systems with relays over Rayleigh-fading channels," \emph{IEEE
Trans. Wireless Communications}, vol. 2, no. 6, pp. 1126-1131, Nov.
2003.

\bibitem{Finite_Mixture_Models}
G. Mclachlan, D. Peel, \emph{Finite Mixture Models}.  New York:
Wiley, 2000.
\bibitem{Computional_Statistic_Matlab}
 W. Martinez, A. Chapman, \emph{Computational Statistics Handbook with
 MATLAB}. Florida: Chapman and Hall/CRC, 2002.

\bibitem{Fading_Alouni}
 S. Simon, M. Alouini, \emph{Digital Communication over Fading
 Channels}. New Jersy: John Wiley \& Sons Inc, 2004.
\bibitem{adaptive2}
A. J. Goldsmith and S. G. Chua, ``Variable-rate variable-power MQAM
for fading channel," \textit{IEEE Trans. Commun.}, vol. 45, no. 10,
pp. 1218-1230, Oct. 1997.

\end{thebibliography}
\end{document}